\documentstyle[twocolumn,prl,aps]{revtex}
\title{Universality of low-energy scattering in three-dimensional field theory}
\begin{document}
\input epsf
\input mssymb

\draft
\author{J. Bros and D. Iagolnitzer} 
\address{Service de Physique Th\'eorique, Centre d'Etudes de Saclay, 91191 Gif-sur-Yvette cedex, France}
\maketitle

\noindent
\begin{abstract}
{\bf Abstract.} Universal low-energy behaviour  
${2 m c }\over{ \ln |s-4m^2|}$          
of the scattering function of particles of positive mass $m$  
near the threshold $s=4m^2$, 
and ${\pi} \over {\ln |s-4m^2|}$ for the corresponding $S-$wave phase-shift,  
is established 
for weakly coupled field theory models with mass $m>0$ 
in space-time
dimension 3; $c$ is a numerical constant independent of the model and couplings. 
This result is a non-perturbative property based on an exact analysis of the  
scattering function in terms of 
a two-particle irreducible (or Bethe-Salpeter) structure function. 
It also appears as generic by the same analysis in the framework of general  
relativistic quantum field theory.   

\end{abstract}

\vskip 0.5cm

There has been great interest within the last two decades in two-dimensional
and, more recently, three-dimensional
field theories  
both in view of their conceptual importance and 
for possible physical applications. In this letter, 
we are more specifically concerned with the low-energy behaviour
of the (connected) two-body scattering function $T$ in 
massive field theories in space-time  
dimension $d=2+1;$ for definiteness, we consider
models with one basic physical mass $m$ associated 
with an elementary particle of the theory.
The models rigorously defined so far for $d=3$ include 
$\lambda \phi^4$ and  
$\lambda \phi^4   
+\lambda'  \phi^3 $ at weak couplings [1].    
By low-energy behaviour of $T$, we mean its dominant behaviour  
near the two-particle threshold $s= 4m^2$, which is the lowest
physical value of the squared center-of-mass energy $s$ of the $2 \to 2$ particle 
process considered.
The aim of this letter is:

\noindent

1) to show that, 
for all the previous field models,  
$T$ enjoys a universal low-energy behaviour equal to  
${2mc }\over{ \ln |s-4m^2|}$ (a more complete analysis will be given 
in [2]), and         

\noindent

2) to indicate why such a universal behaviour 
is generic
in the class of all possible $3d-$field theories. 

In fact, this universal behaviour will appear as an intrinsically non-perturbative phenomenon, 
{\it  spontaneously switched-on by the interaction} in (2+1) dimensions.

In order to situate the history of the subject, let us recall the following fact.  
Starting from the basic principles of relativistic quantum field theory (QFT)  
together with unitarity, 
the following result concerning 
the partial waves $f_l(s)$ of the scattering function $T$\  
(defined by $ f_l(s) = {1\over \pi}\int_0^{2 \pi} T(s, cos \theta) \ cos l\theta\ d\theta$)\  
was proven   
\footnote
{Eq.(1) is the special case $d=3$ of Eq (7) of [3b]. 
More precisely, this result follows 
(see in [3b] the argument between Eqs (4) and (7), completed by the note 14)) 
from local physical-sheet
analyticity, proved [4] from the standard L.S.Z. or Wightman
axioms of relativistic quantum field theory [5],
unitarity (written for the partial waves) and a regularity assumption 
(e.g.continuity) on $T$ at $s>4m^2$ in
order to avoid a la Martin pathologies [6].} 
in [3]: 

\vskip 1cm

$$ f_l(s)\ =\ c s^{1\over 2} \ [\ln\sigma\ +\ b_l(s)]^{-1}; \ \ \ \ \ (1)$$   
in the latter, $\sigma = 4m^2 - s ,  
\ b_l$  is real analytic in a neighborhood of $s=4m^2$  
apart from a possible (simple or multiple) pole
at $s=4m^2$, and $c$  is a well-specified constant depending on 
the normalization conventions, but \it not on 
the theory. \rm 

\smallskip 
Considering more specifically  
the $S-$wave $f_0$, the following  
consequence of Eq.(1) has recently been pointed out by  
K.Chadan, N.N.
Khuri, A.Martin and T.T.Wu  
[7]:   

a) either
$b_0$ has no pole at $ s=4m^2$, i.e. is locally analytic and thus bounded, in which 
case $f_0$  behaves near $s=4m^2$  as 
$2 m c\over \ln|\sigma|$ which is a "universal"
behaviour independent of the theory;   
correspondingly, the phase-shift $\delta_0$  
given by $\hat f_0= e^{i\delta_0} sin\delta_0 
= {\pi \over {c \sqrt s} }\ f_0  $ 
behaves as 
$ {\pi} \over \ln|\sigma|$;  

b) or $b_0$ has a pole, e.g. in $1\over \sigma$, in which case $f_0$ 
and $\delta_0$ 
behave instead as   
$cst\  \sigma$ near $s=4m^2$; the constants now depend on the theory. 

\smallskip 
We therefore see that for $f_0$ the derivation of the announced ``universality property'' 
is directly linked to proving that case a) is satisfied
\footnote
{In [7], a rather complete analysis of the same    
structure is given in the framework of non-relativistic scattering theory;   
as an extension of earlier results of [8],  
it is found 
that the universal 
low-energy behaviour   
${\pi} \over {\ln|k|}$\ of $\delta_0$ 
($k$ being the non-relativistic analogue of $\sigma$) 
actually holds  
for a large class of potentials.}  
by all the field models considered 
and is actually generic in the framework of QFT.  
The claim that $b_0(s)$ should have no pole  at $s=4m^2$ is highly non-obvious and non-trivial.
On the contrary 
for $l \ge 1$ 
all $b_l$ do have poles, 
in accord with the fact
(explained later) that 
all the partial waves $f_l, \ l\ge 1$  
and thereby  
the function $T(s,cos \theta) - f_0(s)  
\equiv 2 \sum_{l\ge 1} f_l(s) cosl\theta $  
are bounded by $cst\ \sigma$ near the threshold.  
The proof of the common universal behaviour 
of $T$ and $f_0$ therefore relies on a precise control of $b_0(s)$.
This control will be achieved  
through an analysis 
of field-theoretical structure functions 
{\it which goes beyond perturbation theory.}  
The latter is in fact deeply misleading,     
because all Feynman functions associated with 
two-particle-reducible diagrams (as those of the toy-structure below) 
exhibit divergences at threshold in powers of $\ln \sigma$   
(see [3,7]).     
\smallskip

Our method is based on the existence of an 
{\it exact Bethe-Salpeter equation}, namely
$$ F(K; k,k') \ =\ B(K; k,k')\ + \cdots\ \ \ \ \ \ \ \ $$ 
$$ \int_{\Gamma (K)} 
{F(K; k,{\rm k}) B(K; {\rm k}, k')  
\over{[(K/2 + {\rm k})^2 - m^2] 
[(K/2 - {\rm k})^2 - m^2]}} 
\ \ d^3 {\rm k}\ \ \ (2)$$ 
which allows one to compute
the complete four-point function $F$ of the field, namely the off-shell
extrapolation of the scattering function $T$, in terms of 
a well-defined ``two-particle-irreducible structure function'' 
(or ``Bethe-Salpeter kernel'') called $B$. 
For simplicity, we have written Eq.(2) in such a way that, in their definitions, $B$ and $F$ include
as internal factors the pair of amputated two-point functions 
of the incoming energy-momenta normalized to $1$ on the mass-shell.
\footnote
{This amounts to making a ``wave-function normalization'' which fixes the constant $c$ 
in Eq.(1) via unitarity.}
In Eq.(2), $ K$ represents the total energy-momentum vector of the 
channel considered ($K^2=s$), while $k$ and $k'$ denote
correspondingly the incoming and outgoing relative energy-momenta of this channel; 
these vectors can be complex  
and the (Feynman-type) integral in (2) is taken on a well-specified 
cycle  
$\Gamma (K)$  
in complex ${\rm k}-$space.  
A usual suggestive graphical notation (of generalized Feynman type) for Eq.(2) is
\begin{figure}
\centerline{
\epsfxsize=7truecm 
\epsfbox{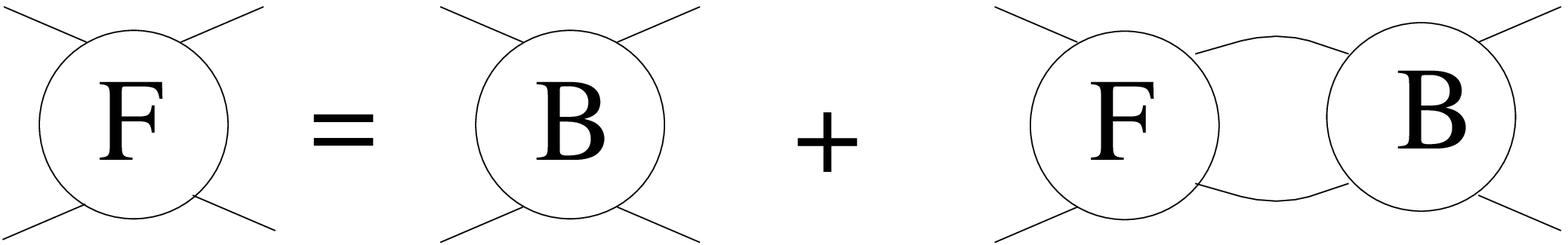}\ \ \ \raise 10pt 
\hbox{(2')}} 
\end{figure}
\noindent
in which the internal lines of the one-loop diagram carry respectively the 
energy-momentum vectors 
$ {\rm k}_1= K/2+{\rm k}$ and 
$ {\rm k}_2= K/2-{\rm k}.$ The mass shell, defined by 
$ {\rm k}_1^2= 
{\rm k}_2^2= m^2$, is (for fixed $K$) a sphere of radius $\sigma$; 
at threshold, the  vanishing of this sphere    
is responsible for the fact that {\it the cycle $\Gamma (K)$ 
becomes pinched by the pair of poles in the 
integral of (2)}.
In a shorter notation we shall also rewrite Eqs (2) or (2')  
as $F = B + F \circ B$, where $\circ$ represents the Feynman-type integration 
of Eq.(2) including two poles. 

In perturbation theory, Eq.(2') just represents a diagram counting identity 
in which the kernel $B$ is  
the (formal) sum of four-point Feynman functions corresponding to all
diagrams which are {\it two-particle-irreducible} 
in the scattering channel considered: {\it only these  perturbative functions  
have no divergences at threshold.} 

The fact that $B$ does exist 
in the exact field theories as an analytic function 
related to $F$ by the Fredholm-type equation (2) 
has been rigorously established in our past works [9], based on the pioneering
ideas of Symanzik [10]. More precisely, two sets of results hold and are  
used here.

\noindent
\ i) Results implied by the basic principles of QFT together with ``off-shell unitarity''
( this is the postulate of ``asymptotic completeness of the fields''); 
they state that:

a) $B$ exists and is analytic in a region
of the $s-$plane ($s=K^2$) containing the threshold $4m^2$: in particular
it has to be {\it uniform} around the threshold [9a,b];   

b) in view of Eq.(2), the corresponding singular structure 
of $F$ at threshold is a pure consequence   
of the pinching of the cycle $\Gamma (K)$ 
by the two poles of the integration operation $\circ$, mentioned above [9a,c]. 

\noindent
\ ii) Results of constructive field theory: for all small coupling
models that have been constructed for $d= 3$ (and $d=2$), 
the function $B$ can be
rigorously defined and controlled in terms of the couplings; 
the property ia) of $B$ is then built-in, 
together with the exact Bethe-Salpeter
equation (2) 
(see [11] 
and our comments below). 

In order to make the ideas of our proof more transparent, it is very illustrative 
to restrict our attention to the $\lambda \phi^4-$model and  
to a ``toy Bethe-Salpeter structure'' already exhibiting all the features of the 
exact Bethe-Salpeter structure. This toy structure is obtained by considering 
the Feynman functions associated with the following bubble diagrams   
\begin{figure}
\epsfxsize=8truecm{\centerline{\epsfbox{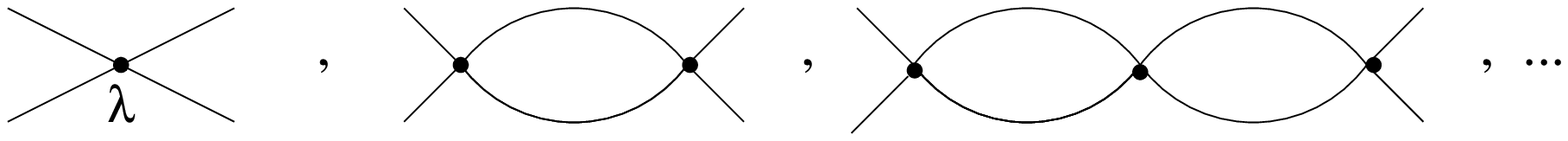}}}
\end{figure}
If $B = B_0^{(\lambda)} \equiv \lambda |1><1|$ denotes the constant kernel equal to 
$\lambda$, represented by the  
single-vertex diagram in the previous picture, 
all these Feynman functions are the successive iterated terms 
$B_0^{(\lambda)}\circ B_0^{(\lambda)} \cdots \circ B_0^{(\lambda)}$ 
(or Neumann series) of a 
Bethe-Salpeter equation $F_0^{(\lambda)} = B_0^{(\lambda)} + F_0^{(\lambda)} \circ B_0^{(\lambda)},$ 
which can be solved by elementary algebra.
In fact, 
$F_0^{(\lambda)}$ is, like $B_0^{(\lambda)}$, a separable (or rank one) kernel given by the formula
$$F_0^{(\lambda)}(s) = {\lambda\ |1><1| \over 1- \lambda <1| \circ |1>(s)},\ \ \ \ (3)$$ 
$$ {\rm where}\ \ \ <1| \circ |1>(K^2)= \cdots$$ 
$$\int { 1 
\over{[(K/2 + {\rm k})^2 - m^2] 
[(K/2 - {\rm k})^2 - m^2]}}\  d^3 {\rm k} =\cdots$$
$$ i\pi^2\  s^{-{1\over2}} \ln \sigma + h(s), \ \ \ h(s)\  {\rm analytic\  at}\  s=4m^2; \ \ (4) $$
$<1| \circ |1>(K^2)$ 
is, up to the factor $\lambda^2$, the one-bubble Feynman function
of the previous picture.
The perturbative expansion of $F_0^{(\lambda)}$, equal to the geometrical series
$$ \sum_{p} \lambda^{p+1} \left[<1| \circ |1>(s)\right]^p=  
 \sum_{p} \lambda^{p+1}       
( cst s^{-{1\over2}} \ln \sigma + h(s))^p, \ \ \  $$
is of course completely misleading since at fixed $\lambda$  
its general ($p$ bubble) term diverges like $(\ln \sigma)^p$ 
at threshold, while in view of (3) and (4)  
the function $F_0^{(\lambda)}$ is in fact a unitary amplitude of the form (1), with
$ b_o(s) = c s^{1\over 2}( h(s)-{1\over \lambda}),\ c=-i\pi^{-2}.$   
For any $\lambda$ different from zero, 
$b_0(s)$ is bounded (like $h(s)$) and therefore 
$F_0^{(\lambda)}$ 
has the universal behaviour $2mc\over \ln \sigma$ near $\sigma=0$.   
 
Coming back to the exact Bethe-Salpeter equation of any given three-dimensional field theory 
with four-point function $F$ and associated scattering function $T$, 
we will now show that the previous toy-structure   
and related  
phenomenons at threshold 
govern basically the corresponding behaviour of the $S-$wave $f_0$ of $T$. 
It is convenient to use the {\it $S-$wave part $F_0$ of the four-point function $F$,} 
whose definition in terms of $F$ 
is similar to that of $f_0$ in terms of $T$ except that it is applied to off-shell Euclidean 
configurations and then analytically continued in the domain implied by the basic principles of QFT. 
$F_0$ is an off-shell extrapolation of $f_0$ which satisfies a partial-wave 
Bethe-Salpeter equation $ F_0 = B_0 + F_0 \circ B_0$ in terms of the $S-$wave part $B_0$ of $B$ [12]. 
The particularity of the kernels $B_0, F_0,\cdots$ with respect to $B,F,\cdots$ is that they only depend
on the Lorentz invariant variables $ K^2=s,\ k_i^2= (K/2\pm k)^2,\   
{k'}_i^2= (K/2\pm k')^2,\ i=1,2,$ and not on the transfer $t = (k-k')^2$. 
The kernels $B_0^{(\lambda)}$ and
$F_0^{(\lambda)}$ of the previous toy-structure are of this nature. 

Our argument is based on the following appropriate splitting $B_0= B'_0 + B''_0$ of the given kernel
$B_0(s; k_1^2,k_2^2; {k'}_1^2, {k'}_2^2)$:   

a) $B'_0 = B_0(s;k_1^2,k_2^2;m^2,m^2)$ is a separable (or rank one) kernel which we write (with the
previous bra-cket notation) $B'_0 = |\Psi_0><1|$ 

b) correspondingly the kernel  
$B''_0(s; k_1^2,k_2^2; {k'}_1^2, {k'}_2^2)$ satisfies the property that   
$B''_0(s;k_1^2,k_2^2;m^2,m^2) \equiv 0$. The same property 
is of course shared by  
all the iterated kernels $B''_0 \circ B''_0\cdots\circ B''_0$ and  
by the corresponding solution $F''_0$ of the auxiliary Bethe-Salpeter  
equation $F''_0 = B''_0 + F''_0 \circ B''_0$; moreover (this is less trivial but crucial),   
all of them are also analytic in $s$  
at $s=4m^2$ up to the ``non-generic case'' of the Fredholm  alternative
which would produce a pole for $F''_0$; in other words {\it no $\ln \sigma$
singularity is present in these kernels}. The reason for it is that $B''_0$  
can be written as a sum of two terms which
respectively factor out ${k'}_i^2 - m^2,\  i= 1,2,$ and therefore cancel out 
either one of the poles in the $\circ $ operation; as an effect, the cycle $\Gamma(K) $ is no more 
pinched at $K^2=4m^2$ and analyticity at threshold is preserved generically in solving this auxiliary
Bethe-Salpeter equation.

The splitting of $B_0$ implies the following relation between the Fredholm resolvents $F_0,F''_0$
of $B_0$ and $B''_0$:
$ {\bf 1} + F_0 = 
({\bf 1} + F''_0)\circ \left[{\bf 1} - B'_0 \circ  
({\bf 1} + F''_0)\right]^{-1}$. 
Taking then into account the special form of $B'_0$ and the fact that $F''_0$ vanishes for 
$ {k'}_1^2 = {k'}_2^2 = m^2$ yields the following exact formula for the 
$S-$wave $f_0(s) = F_0(s;m^2, m^2; m^2,m^2) $ of the theory considered:  
$$ f_0(s) = {(|\Psi><1|)(s;k_1^2,k_2^2)_{|k_1^2=k_2^2=m^2} \over 1\ -\ <1|\circ |\Psi>(s)},\ \ \ (5)$$  
where:
$$|\Psi> = |\Psi_0>\ +\ F''_0 \circ |\Psi_0>.\ \ \ \ \ \ (6)$$
Eq.(5) is of the same form as Eq.(3) , to which it reduces in an obvious way when
$B_0 = B'_0 = B_0^{(\lambda)}$, i.e. when $|\Psi_0> = \lambda |1> $ and $B''_0 = F''_0 = 0.$
However, the remarkable fact is that Eqs (5),(6) are exact (non-perturbative) equations,  
valid for the $S-$wave of any field theory.  
We shall now show that Eq.(5) exhibits a structure which is exactly of the form (1)  
with $b_0(s)$ bounded in ``generic cases'' 
and in particular for all weakly-coupled models. 

We notice that the function $\Psi(s;k_1^2,k_2^2)$ represented by $|\Psi>$ (or by the kernel 
$|\Psi><1|$) is generically analytic at $s=4m^2$: this follows from  
Eq(6) by applying to $F''_0$ the ``non-pinching argument'' given above in b).  
Let us now call $g(s)$ the numerator at the r.h.s. of Eq.(5), namely 
$g(s) = \Psi(s; m^2,m^2)$. Then we claim that one can write:
$$ <1|\circ|\Psi>(s) = <1|\circ|1>(s)\times g(s) + l(s),\ \ \ \ (7)$$
where $l(s)$ is generically analytic at $s=4m^2$. This follows from 
writing  
$\Psi(s;k_1^2,k_2^2) = g(s) + \Psi_{(reg)}(s; k_1^2,k_2^2),$ where
$\Psi_{(reg)}$ vanishes at $k_1^2 = k_2^2 =m^2$ and therefore produces
a function $l(s) = <1|\circ|\Psi_{(reg)}>$ regular at threshold 
(again in view of the non-pinching argument of b), but used on the left).  
In view of (7) (and by taking (4) into account), we can thus rewrite Eq.(5) 
as follows:
$$ f_0(s) = {g(s) \over 1-l(s) -g(s) (c^{-1} s^{-{1\over2}} \ln \sigma + h(s))}  
\ \ \ \ \ (8).$$   
We then conclude that $f_0$ is of the form (1), with 
$$ b_0(s) = c s^{1\over2} \left( h(s) + {l(s)-1 \over g(s)} \right) .\ \ \ \ (9)$$ 
Since $h$ is analytic, the question of the universality of $f_0$ in $(\ln \sigma)^{-1}$ 
amounts to discussing the generic character of the fact that $(l(s)-1)/g(s)$ has no 
pole at $s=4m^2$. The situation is as follows:

i) In all weakly-coupled models containing a $\lambda \phi^4-$term,  
it is claimed (see our comment below) that 
$B$ is of the form $B= B_0^{(\lambda)} + O(\lambda^2)$.  
This entails that  
$\Psi_0 = \lambda + O(\lambda^2)[s;k_1^2,k_2^2]$, while $B''_0$ and therefore $F''_0$ 
are bounded analytic functions of  
$s,k_1^2,k_2^2,{k'}_1^2, {k'}_2^2$ of order   
$O(\lambda^2)$. It then follows from Eq.(6) that    
$\Psi = \lambda + O(\lambda^2)[s;k_1^2,k_2^2]$ and therefore  
$g(s) = \lambda + O(\lambda^2)[s]$,   
while the analytic functions $\Psi_{(reg)}(s;k_1^2,k_2^2)$ and therefore  
$l(s)$ are of order  
$O(\lambda^2)$. One thus concludes that      
$${l(s)-1 \over g(s)}=  
{O(\lambda^2)[s]-1\over   
\lambda + O(\lambda^2)[s]},\ \ \ \ \ (10)$$
which behaves like $-{1\over \lambda}$ 
and is therefore {\it finite} in some range of couplings ($0< \lambda < \lambda_0$):
for this class of theories the universal behaviour of $f_0$ at threshold is thus
established; moreover, one can see as a by-product that  
$f_0$ exactly behaves as the function $F_0^{(\lambda)}$ of the toy-structure in the 
limit of small $\lambda'$s.

ii) For more general field theories, the universal behaviour of $f_0$ is valid
except if either $g(s)=0$ or $l(s) = \infty$, the latter case being produced by the Fredholm
alternative in the auxiliary Bethe-Salpeter equation (i.e. $F''_0 =\infty$).
These exceptions are defined by the 
vanishing of analytic functions which in view of i) cannot be identically zero 
(at least under the usual postulate of analytic continuation in the couplings). Such cases necessitate  
that $B''_0$ is large i.e. that $B_0$ has a large rate of variation
with respect to the masses near the mass-shell.

\smallskip
A similar analysis can be made for $T$ 
(including in the models the dependence  
of $T$ with respect to the couplings near $s=4m^2$) [2]. However,  
the fact that $T$ and $f_o$ enjoy the same universal behaviour at threshold 
(with the possible exceptions analysed above) is implied by the following general property:
$T-f_0$ is bounded at small $\sigma$ by $cst\ {|\sigma|}$.  
This property relies on the analyticity of $\hat T(s,t) \equiv T(s,cos \theta)$ in a \it fixed \rm 
neighborhood of $t=0$ in the complex plane of the variable $t={{4m^2-s}\over 2}(1-cos\theta)$,
for all $s$ in a cut-neighborhood of $4m^2$,   
which is a rigorous result [4] of general QFT (under a regularity assumption  
of the type specified in [6], automatically satisfied in the models).  
The latter property implies that, if one rewrites 
$T-f_0 = \sum_{l \ge 1} f_l(s)(z^l + z^{-l})$ with $z= e^{i \theta}$,
this series in $z$
has to converge in a ring of the form $ cst\ \sigma < |z| < (cst\ \sigma)^{-1}$,  
where it is uniformly bounded in $\sigma$.
By a standard argument of complex analysis, one concludes that    
each $f_l(s)$ is bounded at small $\sigma$ by $cst\ {|\sigma|}^l$ 
which then entails the announced bound on $T-f_0$.   
(These bounds on the $f_l$ imply that for $l \ge 1$,    
each uniform function $b_l$ of Eq.(1) does have a pole \it of order \rm $\ge l$.) 

\vskip 0.2cm
\noindent
{\sl Results of constructive QFT for weakly-coupled $3d-$field models:} 

The above mentioned results on $B$ can be obtained 
by combining a rigorous definition of $3d$-models of the type given for the model 
$\lambda \phi^4$ in [1] with a definition of $B$ of the type given in [11a] for 
two-dimensional models, which is easily adaptable to the $3d-$case. They can also 
be derived as a specially simple case through the general methods given in [11b] for 
treating non-superrenormalisable models (such as the massive Gross-Neveu model in dimension 2).  
In these methods of constructive QFT, one defines the Fourier transforms 
$\tilde F$ and $\tilde B$ of $F$ and $B$ in Euclidean space-time through a certain type 
of expansions which are convergent at small couplings, in contrast to perturbative expansions. 
Diagrammatical analysis provides Eq.(2) (or (2')). On the other hand, expansions of $\tilde B$
only involve two-particle irreducible diagrams and this fact implies that $\tilde B$ satisfies better
exponential fall-off properties than $\tilde F$ (in Euclidean space-time). 
This implies in turn that $B$ is analytic in a larger strip around Euclidean energy-momentum space,  
which includes the threshold $s=4m^2$. 
The analysis distinguishes the single vertex part equal to the constant $\lambda$ from a remainder
(i.e. the sum of all the other two-particle irreducible contributions) that is uniformly bounded
by $cst\ \lambda^2$ in the whole strip of analyticity. Details will be given elsewhere [2b].

\vskip 0.2cm
\noindent
{\sl Final remarks:}

i) In the models, the universal behaviour is valid in a region around 
$s=4m^2$ which shrinks to zero as $\lambda \to 0$ (in exp[-cst/$\lambda$] in view of Eqs (8),(9),(10)):
it is thus consistent with the fact that $ T \equiv 0$ for the free field. 
In other words, universality appears as soon as the coupling is switched-on.  

ii) All the present results can be alternatively obtained by another method (presented in [2]
and already applied in the past to the related problem of bound states [13a]) 
which exhibits other interesting features. 
This method is based on the use of another type of  
irreducible kernel $U$ (derived from $B$ through an integral
equation without singularity at threshold [13b]) 
which is the analogue for $d=3$ of Zimmermann's $K-$matrix [14].

\vskip 0.5cm
\noindent
[1] J.Magnen and R.Seneor, Ann. Inst. Henri Poincar\'e {\bf 24}, 95 (1976) and references therein. 

\noindent
[2]a) J. Bros and D. Iagolnitzer, ``Physics near threshold in space-time dimension 3'';
b) D.Iagolnitzer and J. Magnen, in preparation.

\noindent
[3] J. Bros and D. Iagolnitzer, a) Commun. Math. Phys. {\bf 85}, 197 (1982),  
b) Phys. Rev. D {\bf 27}, 811 (1983).

\noindent
[4] J. Bros, H. Epstein and V. Glaser, Nuovo Cimento {\bf 31}, 1265 (1964).  

\noindent
[5] R. F. Streater and A. S. Wightman, \it PCT, Spin and Statistics, and All That \rm  
(W. A. Benjamin, New York, 1964); 
H. Lehmann, K. Symanzik and W. Zimmermann, Nuovo Cimento {\bf 1}, 205 (1955) and 
{\bf 6}, 319 (1957).  

\noindent
[6] A. Martin, in \it Problems in Theoretical Physics, \rm D.I. Blokintseff et al. eds  
(Nauka, Moscow, 1969), p. 113; 


\noindent
[7] K. Chadan, N. Khuri, A. Martin and T. T. Wu, 
Phys. Rev. D {\bf 58}, 025014 (1998).

\noindent
[8] D. Boll\'e and F. Gesztesy, Phys. Rev. Lett. {\bf 52}, 1469 (1984); 
P. G. Averbuch, J. Phys. A  {\bf 19}, 2325 (1986);  

\noindent
[9]a) J. Bros, in \it Analytic Methods in Mathematical Physics \rm (Gordon and Breach, New York, 1970),
pp. 85-135; b) J. Bros and M. Lassalle, Commun. Math. Phys. {\bf 54}, 33 (1977); 
c) J. Bros and D. Pesenti, J. Math pures et appl. {\bf 58}, 375 (1980) and 
{\bf 62}, 215 (1983).   

\noindent
[10] K. Symanzik, J. Math. Phys. {\bf 1}, 249 (1960). 

\noindent
[11] D. Iagolnitzer and J. Magnen, Commun. Math. Phys. a) {\bf 110}, 51 (1987) and 
b) {\bf 111}, 81 (1987).  

\noindent 
[12] J.Bros and G.A. Viano, in \it Rigorous Methods in Particle Physics \rm, 
S. Ciulli, F. Scheck and W. Thirring, eds. (Springer tracts in Mod. Phys. {\bf 19}, 53 (1990)).  

\noindent
[13]a) J. Bros and D. Iagolnitzer, Commun. Math. Phys. {\bf 119}, 331 (1988);   
b) D. Iagolnitzer, Commun. Math. Phys. {\bf 88}, 235 (1983).   

\noindent
[14]  W. Zimmermann (1961), Nuovo Cimento {\bf 21}, 429 (1961) and references therein. 

\end{document}